\documentclass[aps,prd,twocolumn,amsmath,amssymb,amsfonts,nofootinbib,superscriptaddress,altaffilletter]{revtex4-1}
\usepackage{graphicx}
\usepackage{hyperref}
\hypersetup{
  bookmarksopen=true
}
\usepackage{amssymb}
\usepackage{amsmath}
\usepackage{color}
\usepackage{supertabular}
\usepackage{dcolumn}

\def\ec{\textrm{~,}}

\def\fdot{f_0^{(1)}}

\def\etal{{\it et al.}}

\newcommand{\Fstat}{$\mathcal{F}$-statistic}

\def\sci#1#2{#1\times10^{#2}}

\def\RAJ{\textrm{RA}_{\textrm J2000}}
\def\DECJ{\textrm{DEC}_{\textrm J2000}}

\begin{document}

\title{ 
Loosely coherent search in LIGO O1 data for continuous gravitational waves from Terzan 5 and the galactic center
}

\author{Vladimir Dergachev}
\email{vladimir.dergachev@aei.mpg.de}
\affiliation{Max Planck Institute for Gravitational Physics (Albert Einstein Institute), Callinstrasse 38, 30167 Hannover, Germany}
\affiliation{Leibniz Universit\"at Hannover, D-30167 Hannover, Germany}

\author{Maria Alessandra Papa}
\email{maria.alessandra.papa@aei.mpg.de}
\affiliation{Max Planck Institute for Gravitational Physics (Albert Einstein Institute), Callinstrasse 38, 30167 Hannover, Germany}
\affiliation{Leibniz Universit\"at Hannover, D-30167 Hannover, Germany}
\affiliation{University of Wisconsin Milwaukee, 3135 N Maryland Ave, Milwaukee, WI 53211, USA}

\author{Benjamin Steltner}
\email{benjamin.steltner@aei.mpg.de}
\affiliation{Max Planck Institute for Gravitational Physics (Albert Einstein Institute), Callinstrasse 38, 30167 Hannover, Germany}
\affiliation{Leibniz Universit\"at Hannover, D-30167 Hannover, Germany}

\author{Heinz-Bernd Eggenstein}
\email{heinz-bernd.eggenstein@aei.mpg.de}
\affiliation{Max Planck Institute for Gravitational Physics (Albert Einstein Institute), Callinstrasse 38, 30167 Hannover, Germany}
\affiliation{Leibniz Universit\"at Hannover, D-30167 Hannover, Germany}


\begin{abstract}
  We report results of a search for continuous gravitational waves from a region covering the globular cluster Terzan 5 and the galactic center.  Continuous gravitational waves are expected from fast-spinning, slightly non-axisymmetric isolated neutron stars as well as more exotic objects. The regions that we target are believed to be unusually abundant in neutron stars. We use a new loosely coherent search method that allows to reach unprecedented levels of sensitivity for this type of search. The search covers the frequency band 475--1500\,Hz and frequency time derivatives in the range of $\sci{[-3.0, +0.1]}{-8}$\,Hz/s, which is a parameter range not explored before with the depth reached by this search.   
  As to be expected with only a few months of data from the same observing run, it is very difficult to make a confident detection of a continuous signal over such a large parameter space. A list of parameter space points that passed all the thresholds of this search is provided. We follow-up the most significant outlier on the newly released O2 data and cannot confirm it.
We provide upper limits on the gravitational wave strength of signals as a function of signal frequency. 
\end{abstract}

%
%
\maketitle

\section{Introduction}
\label{sec:introduction}


Continuous gravitational waves (CWs) are expected from fast-spinning neutron stars in a variety of circumstances, for example if they present a slight non-axisymmetry (ellipticity). Many CW searches have been carried out on LIGO 
data~\cite{rilesmpla}, including several all-sky searches ~\cite{O1EH,O1AllSky1,O1AllSky2, o1_allsky3} 
and broadband directed searches~\cite{orionspur}. No signals have been detected yet.

Directed searches are searches for signals from interesting targets -- both specific objects or/and regions.  
The search presented here, targeting emission from the globular cluster Terzan 5 and the galactic center,  falls into this category. 

We use data 
collected during the first Advanced LIGO observing run, O1, ~\cite{aligo, O1DetectorPaper, o1_data, losc} and employ a new medium-scale loosely coherent algorithm \cite{loosely_coherent,loosely_coherent2,loosely_coherent3}. We probe a broad class of signals with frequencies between 475 and 1500 Hz, with unprecedented sensitivity. For sources at 8.5 kpc this search is sensitive to signals from neutron star deformations well within the range allowed by conventional neutron star equations of state~\cite{McDanielJohnsonOwen}.  

Additionally this search was used as a pilot study of the new loosely-coherent search method. The search uses a substantially longer coherence length than used before and hence presents most of the challenges and difficulties of an all-sky search, but without the substantial load of searching the whole sky. This search has exposed performance bottlenecks in the algorithms implementation and has paved the way for the first all-sky loosely coherent search \cite{o1_allsky3}.

The paper is organized as follows: sections \ref{sec:LIGO-O1} and \ref{sec:SignalWaveform} briefly introduce the LIGO detectors, the data that is used and the signal waveform that we target with this search. Section \ref{sec:looselyCoherent} describes the  features of the main building block of the search, the enhanced loosely coherent method, and section \ref{sec:pipeline} illustrates the pipeline, including the way the upper limits are established and the ranking of the outliers. The results are presented and discussed in section \ref{sec:results}. The appendix \ref{sec:outliers} contains the outlier tables.


\begin{figure*}[htbp]
\begin{center}
  \includegraphics[width=7.2in]{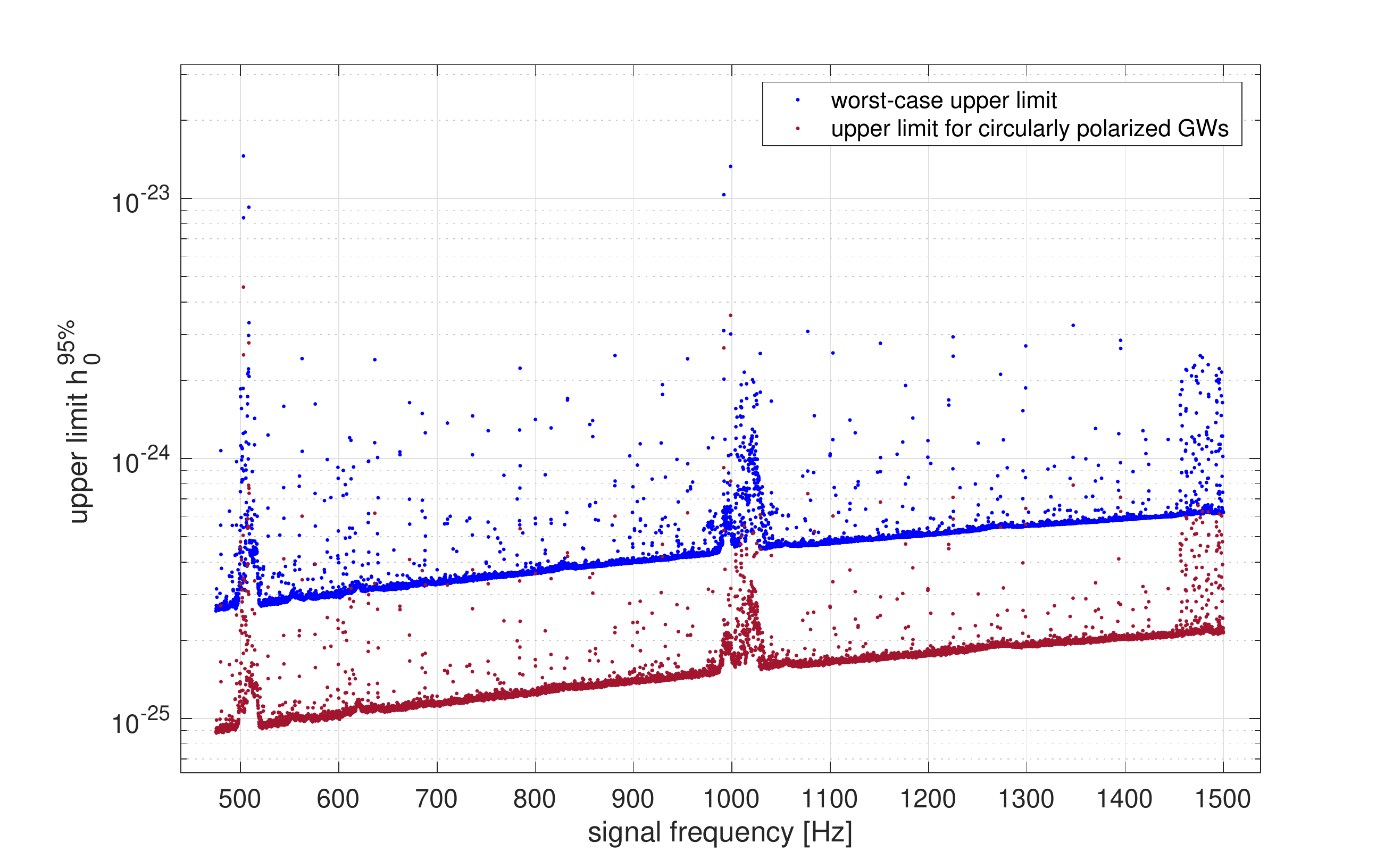}
 \caption{Upper limits on gravitational wave strain. The dimensionless strain $h_0$ (vertical axis) is plotted against signal frequency. The upper (blue) curve shows worst-case (linearly polarized) $95$\% confidence level (CL) upper limits as a function of frequency. The upper limits are maximized over sky and all intrinsic signal parameters for each frequency band. The lower (red) curve shows upper limits assuming a circularly polarized source.  The data for this plot can be found in \cite{data}.
(color online)}
\label{fig:O1_upper_limits}
\end{center}
\end{figure*}

\section{LIGO interferometers and the O1 observing run}
\label{sec:LIGO-O1}

The LIGO gravitational wave detector consists of two 4\,km dual-recycling Michelson interferometers, one in Hanford, Washington and the other in Livingston, Louisiana, separated by a 3000-km baseline.
The interferometer mirrors act as test masses, and the passage of a gravitational wave induces a differential arm length change that is proportional to the gravitational-wave strain amplitude.
The Advanced LIGO~\cite{aligo,O1DetectorPaper} interferometers came online in September 2015 after a major upgrade. 

The O1 run occurred between September 12, 2015 and
January 19, 2016, from which approximately 77 days and 66 days of analyzable
data were produced by the Hanford (H1) and Livingston (L1) interferometers, respectively.

Notable instrumental contaminants affecting the searches described here
include spectral combs of narrow lines in both interferometers, many of
which were identified after the run had ended and were mitigated for future runs~\cite{O1AllSky1, O1AllSky2, Covasetal}.
For instance an 8-Hz comb in H1 with the even harmonics (16-Hz comb)
being especially strong, was ascribed to digitization roundoff
error in a high-frequency excitation applied in order to servo-control the cavity length
of the Output Mode Cleaner (OMC). Similarly, a set of lines found to be linear
combinations of 22.7 Hz and 25.6 Hz in the L1 data was tracked down to digitization error in an OMC
excitation at a still higher frequency.


Although most of these strong and narrow lines are stationary in frequency
and hence do not exhibit the Doppler modulations due to the Earth's motion
expected for a CW signal from most sky locations, they do degrade the sensitivity 
to astrophysical signals at the frequencies where they occur. 


\section{Signal waveform}
\label{sec:SignalWaveform}

In this paper we assume a standard model of a spinning non-axisymmetric neutron star. Such a neutron star radiates circularly-polarized gravitational radiation along the rotation axis and linearly-polarized radiation in the directions perpendicular to the rotation axis. For the purposes of detection and establishing upper limits the linear polarization is the worst case, as such signals contribute the smallest amount of power to the detector.

The strain signal measured by a detector is  
\begin{equation}
\begin{array}{l}
h(t)=h_0\left(F_+(t, \alpha_0, \delta_0, \psi)\frac{1+\cos^2(\iota)}{2}\cos(\Phi(t))+\right.\\
\quad\quad\quad \left.\vphantom{\frac{1+\cos^2(\iota)}{2}}+F_\times(t, \alpha_0, \delta_0, \psi)\cos(\iota)\sin(\Phi(t))\right)\ec
\end{array}
\label{eq:h_of_t}
\end{equation}
where $F_+$ and $F_\times$ are the detector responses to signals with ``$+$'' and ``$\times$'' 
quadrupolar polarizations \cite{S4IncoherentPaper, EarlyS5Paper, FullS5Semicoherent}, the sky location of the source is described by right ascension $\alpha_0$ and declination $\delta_0$, the inclination of the source rotation axis to the line of sight is  $\iota$, and
we use $\psi$ to denote the polarization angle (i.e. the projected source rotation axis in the sky plane).

The phase evolution of the signal is given by
\begin{equation}
\label{eqn:phase_evolution}
\Phi(t)=2\pi\left(f_0\cdot (t-t_0)+\fdot\cdot (t-t_0)^2/2\right)+\phi\ec
\end{equation}
with $f_0$ being the source frequency and $\fdot$ denoting the first frequency derivative (which, when negative, is termed the {\em spindown}). 
We use $t$ to denote the time in the Solar System barycenter frame. The initial phase
$\phi$ is computed relative to reference time $t_0$.  When expressed as a function of local time of ground-based detectors, Equation \ref{eqn:phase_evolution} acquires  sky-position-dependent Doppler shift terms.

Most natural ``isolated'' sources are expected to have negative first frequency derivative, due to the energy lost to emission of gravitational or electromagnetic radiation. The frequency derivative can be positive because of residual motions due to, for instance, a long-period orbit.

\section{The medium scale loosely coherent search}
\label{sec:looselyCoherent}

The medium scale loosely coherent search is the basic building-block of this search. It is described in \cite{loosely_coherent3} and follows earlier loosely coherent implementations \cite{loosely_coherent, loosely_coherent2}.  Here we highlight features that are useful to understand search output, in particular upper limits and outliers.

%
The input to the search are Hann-windowed 3600\,s short Fourier transforms (SFTs) for each of the LIGO interferometers : $\left\{a_{tfi}\right\}$, indexed by time $t$, discrete frequencies $f$ and interferometer index $i$. A value of the weighted power sum $P(f_0,\vec{p})$ is computed for every searched wave shape, parametrized by the frequency of the source $f_0$ and a set of values for its spindown, sky position and source orientation $\vec{p}=(\alpha,\delta,\fdot,\iota)$.

The loosely coherent weighted power sum is a bilinear function of the SFT data:
\begin{equation}
\label{eq:P}
P(f_0,\vec{p})=\frac{\sum_{t_1, t_2, i_1, i_2}K(t_1, t_2, \vec{p}, f_0)a_{t_1 f'_1 i_1}\bar{a}_{t_2 f'_2 i_2}}{\sum_{t_1, t_2}W(t_1, t_2, \vec{p})}.
\end{equation}
Here $f'_1$ and $f'_2$ are the interferometer-frame signal frequencies at the detector-time $t_1$ and $t_2$. 
The kernel $K(t_1, t_2, \vec{p}, f_0)$ is equivalent to a narrow band filter on the input data that includes phase corrections to account for the signals' Doppler shifts and relativistic effects. The weight term $W(t_1, t_2, \vec{p})$  folds-in the noise level of the individual SFTs and the detectors' response to the specific source as a function of time (it is fourth order in the antenna response). The explicit expressions for these functions are very involved not very illuminating without extensive additional information.  We hence do not report them here but rather refer the interested reader to sections II\,-\,IV of \cite{loosely_coherent3}.


Because the polarization coefficients are factored out of power sums (Eq. \ref{eq:P}), which involve thousands of SFTs, it is easy to produce separate power sums for any polarization of interest. For instance, we will provide upper limits for a population of circularly polarized signals which corresponds to the star's rotation axis pointing towards us ($\iota = 0$ or $\pi$ in Eq. \ref{eq:h_of_t}). 

The fact that we compute power sums makes it possible to set upper limits on the signal strain amplitude by estimating the power excess that we would measure from the target signals at a given strain amplitude. This estimate is computed using the  universal statistics algorithm which produces statistically valid results without assumptions on the probability distribution function of the noise -- a rigorous derivation of the algorithm is given in \cite{universal_statistics}. An intuitive explanation of why this is possible is that if the expected power of the noise is bounded, then the expectation of any continuous function of the noise is also bounded over the space of all probability distributions (in mathematical terminology the space of probability distributions is compact in weak topology). If the noise is Gaussian, the implementation of the {\em Universal} statistic used in this search provides close-to-optimal values.

In order to bracket the range of upper limit strain values, depending on the orientation of the source, we consider the so called ``worst-'' and ``best-" case polarization upper limits. The upper limits are given as a function of frequency and apply to 0.125 Hz signal-frequency intervals, i.e. there is a single upper limit number for every 0.125 Hz band. The ``worst-case'' upper limits are based on the maximum universal statistic value over the frequencies in any given band and all spindowns, sky positions and polarizations, further increased (by 7\%) to account for losses due to signal-template mismatch\footnote{The 7\% is derived from the results of Monte Carlo simulations of this search on simulated signals\cite{loosely_coherent3}.}. This maximization tends to select increased universal statistic values due to disturbances in the data, when present.  
For this reason the worst-case upper limit curve has larger outliers than the circular polarization (``best-case'' one).  The ``best-case'' upper limits are based on the maximum universal statistic value over the frequencies in any given the band and all spindowns and sky positions, while circular polarization is assumed. 

The computation of universal statistic \cite{universal_statistics} also computes SNR as a byproduct, this is used as a detection statistic for identifying outliers.

\section{Search pipeline}
\label{sec:pipeline}

We search a disk on the sky of radius 0.06\,rad ($3.43^\circ$) centered on right ascension 4.65\,rad ($266.42^\circ$) and declination $-$0.46\,rad ($-26.35^\circ$). This search area is chosen to cover both the globular cluster Terzan 5 and Sagittarius A$^*$, galactic regions expected to contain many neutron stars. Terzan 5, in particular, has many known radio pulsars \cite{TerzanPulsarPaper1,TerzanPulsarPaper2,ATNF}.

\begin{table}[htbp]
\begin{center}
\begin{tabular}{rD{.}{.}{2}D{.}{.}{3}}\hline
Stage & \multicolumn{1}{c}{Coherence length (hours)} & \multicolumn{1}{c}{Minimum SNR}\\
\hline
\hline
0  & 8  & 6 \\
1  & 12 & 6.5 \\
2  & 16 & 7 \\
3  & 24 & 8 \\
4  & 36 & 9 \\
5  & 48 & 11 \\
6  & 72 & 13 \\
\hline
\end{tabular}
\end{center}
\caption{Search pipeline}{Parameters of search pipeline. As explained in the text stage 6 also features an additional consistency check between the single-detector statistics.}
\label{tab:pipeline_parameters}
\end{table}

The search pipeline iteratively uses the medium scale loosely coherent algorithm in a cascade of 7 different stages. The first stage employs an 8~hour coherence length.  Outliers identified at this stage are followed-up with more sensitive searches that utilise increasingly longer coherence lengths, as detailed in Table \ref{tab:pipeline_parameters}. For all stages the detection statistic combines coherently over the coherent length the data from both detectors. At the last stage, the detection statistic from each detector separately is also computed and the additional requirement is set on surviving candidates that the parameters be consistent across the multi-detector and single-detector statistics.
The consistency condition demands that outliers from the same sky point and spindown are no further than $5$\,$\mu$Hz in frequency.

The pipeline is validated using extensive Monte Carlos that simulate signals in the real data and test the recovery efficiency of the whole pipeline. This approach is completely standard for this type of search, where the expected signals are weak and in many frequency bands it is impossible to model the noise reliably. This procedure also validates the correctness of the upper limit values given here.

\subsection{Outlier ranking}
\label{sec:GFA}
The likelihood of a search outlier to have astrophysical origin is commonly described by the false alarm rate - an estimate of probability that this outlier is produced by pure chance. The most obvious method of computing this rate is to repeat the search many times with different realizations of the noise and count how many similar outliers are produced. This is impractical for broad parameter searches which usually take weeks to months to complete.

A commonly used shortcut is to reuse the data from the original search but combine it differently, for instance with non-astrophysical offsets for coincidence parameters (such as time or frequency) -- for a notable example see \cite{GW150914}. The idea is to simulate different noise realisations of the search results, by constructing ``off-source'' combinations of the actual search results. Unfortunately, producing an ``off-source'' noise realization by combining the single-detector outliers from the last stage of this pipeline is not viable because the preceding stages are based on multi-detector statistics. This means that the outliers at the last stage present correlations between the frequencies of peaks in single-detector data.  We want the artificially generated noise realisations (the off-source data) to also display such correlations. Unfortunately the standard methods to construct the off-source data by recombining the single-detector candidates with non-physical offsets would destroy such correlations, hence they are not suitable.

We take here a different approach and derive an approximate analytical expression, under the assumption that underlying noise is Gaussian. This is a strong assumption that is known not to hold in many frequency bands. Thus this expression should not be used as criterion for detection. Rather it is meant as a figure of merit to evaluate relative significance of outliers.

As the entire hierarchical  7-stage pipeline is difficult to model, we derive the false alarm rate for a hypothetical search that used the last stage of followup to analyze the entire parameter space. In the next paragraphs we describe the quantities that are necessary in order to estimate the false alarm rate Eq. \ref{eq:logGFA}. These quantities are: the total number of templates $N$ that would have been used by the stage 6 search over the entire searched parameter space; the distribution of the detection statistic for the stage 6 search, $P_{\chi^2, k}$; the ``coincidence probability'' associated with the multi-detector/single-detector consistency check, $p_{\textrm{coinc}}$. We derive these below.

We (over)-estimate the total number of templates $N$ necessary to perform such search to be $\sci{1.6}{27}$. We arrive at this number as follows: The total number of templates in the grid for the entire search over 1025 Hz, the whole sky, polarization and spindown  is $\sci{9.3}{21}$. We however search more waveforms than these because we additionally allow the frequency to 
change by up to one frequency bin 11 times, equally spaced throughout the observation period. This adds robustness to our search with respect to deviations of the real signal from a strictly coherent signal model. 
To account for this, we increase $\sci{9.3}{21}$ by a factor of $3^{11}$. This overcounts the number of independent templates. For example, two templates different only by a single jump in frequency bin in the middle of the run, would be highly correlated.

Because we consider the last stage as a separate search the frequencies of outliers in individual interferometers are independent. The frequency coincidence criterion can be falsely triggered in pure noise with probability $p_{\textrm coinc}=\sci{3.59}{-5}$.


The last stage of the analysis uses a 3-day coherence time. As the variations in $W$ (Eq.~\ref{eq:P}) due to amplitude modulations average out over this time, 
the power sums can be modelled as a $\chi^2$ variable 
with at most $k=80$ degrees of freedom, with $k$ expected to  be smaller for frequency regions with highly contaminated data. The reason for decrease in $k$ is that the terms in the sum (Equation \ref{eq:P}) containing contaminated data are de-weighted and hence they contribute less than others to the total number of degrees of freedom. In the case of equal weighted data $k=80$ because there are 40 3-day chunks in a 4-month run and each chunk contributes two degrees of freedom.

We take the Gaussian false alarm figure of merit for a candidate at signal-to-noise ratio value SNR, at the end of the last follow-up stage, to be
\begin{equation}
\begin{array}{l}
\log_{10}\left(\textrm{GFA}\right)=\log_{10}\left(P_{\chi^2, k}\left( k+\sqrt{2k}\cdot\textrm{SNR}\right)\right)+\\
\qquad\qquad+\log_{10}\left(N\right)+\log_{10}\left(p_{\textrm{coinc}}\right),
\end{array}
\label{eq:logGFA}
\end{equation}
where SNR is defined as the ratio of the deviation of the detection statistic from its expected value to the standard deviation. 

We emphasize again that the formula \ref{eq:logGFA} was derived under the assumption of stationary Gaussian noise that is independent between the H1 and L1 interferometers. Since this assumption is violated in many frequency bands, this figure is not meant as a criteria for detection. For example, large negative values for outliers 1 through 8 are an indication of a presence of a signal, but these signals is known to be instrumental in origin.



\section{Results}
\label{sec:results}

The search produces a number of outliers, the strongest of which are traced to clear instrumental artifacts. A number of unclassified outliers with smaller signal-to-noise ratios passes the follow-up pipeline. While the pipeline has been demonstrated to recover injected signals successfully even in the most heavily contaminated regions \cite{loosely_coherent3}, the presence of noise does increase the false alarm rate. As the O1 data is highly contaminated with both stationary and non-stationary instrumental lines, classification of weak outliers is particularly difficult. This problem is made more challenging by the presence of instrumental artifacts coherent between both interferometers. 

We further extend the coherent baseline of the search with ad-hoc semi-coherent follow-up searches like the ones used in \cite{O1AllSky1,O1EH}, on 352 outliers. We use three stages with coherent
baselines of $210$ hrs (12 segments), $500$ hrs (6 segments) and $1260$ hrs (2 segments), respectively.  We denote the stages by FU0, FU1 and FU2. Since FU1 is rather computationally intensive we
do not follow-up any outlier that can be associated with a disturbance (see comment field in the tables of Appendix \ref{sec:outliers}. 21 outliers survive all thresholds from these follow-up
searches. The outlier with id 68 appears to be the most significant. On it we perform a dedicated search using the FU1 search on $480$ hrs of the newly released data from the O2 run
\cite{o1_data}. The search could not recover the candidate with detection statistic values consistent with what would have been expected if outlier 68 had been a continuous wave described by Eq.
\ref{eq:h_of_t}. Appendix \ref{sec:outliers} details all the outliers and indicates at what stage of these follow-ups the candidate was rejected. 

The simulations described in \cite{loosely_coherent3} have shown that an astrophysical source adhering to expected signal model will be recovered within $15$\,$\mu$Hz of true frequency and within $\sci{1.5}{-11}$\,Hz/s of true spindown. The sky position mismatch depends on frequency and, for outliers with frequency $f$ is no more than $\sci{6.5}{-4} \cdot \left(1\,\textrm{kHz}/f\right)$ in ecliptic distance, defined as the distance between outlier location and true injection location after projection onto the ecliptic plane.

The universal statistic algorithm allows to set valid upper limits even in the most heavily contaminated bands. Figure \ref{fig:O1_upper_limits} shows the best-case and worst-case 95\% confidence upper limits on the signal strain in 0.125 Hz frequency intervals. 
At the highest frequency (1500\,Hz) the worst-case upper limit on gravitational wave strain is $\sci{6.2}{-25}$, which translates in a source with an ellipticity of $\sci{2.5}{-6}$ at 8.5 kpc.  Because of maximization procedure the confidence level of the worst-case upper limits remains 95\% or higher for any subset of parameters. For example, if one picks a sky location of the Terzan 5 globular cluster, spin-down of $\sci{5}{-9}$\,Hz/s and a frequency of 550\,Hz the worst case upper limit is $\sci{2.89}{-25}$\,Hz , with a confidence level which is guaranteed to be at least 95\%. The actual confidence level is likely to be larger than 95\% for the specific point, because the quoted upper limit is the highest over all sampled spin-downs and the wider sky area.

\begin{figure}[htbp]
\includegraphics[width=3.5in]{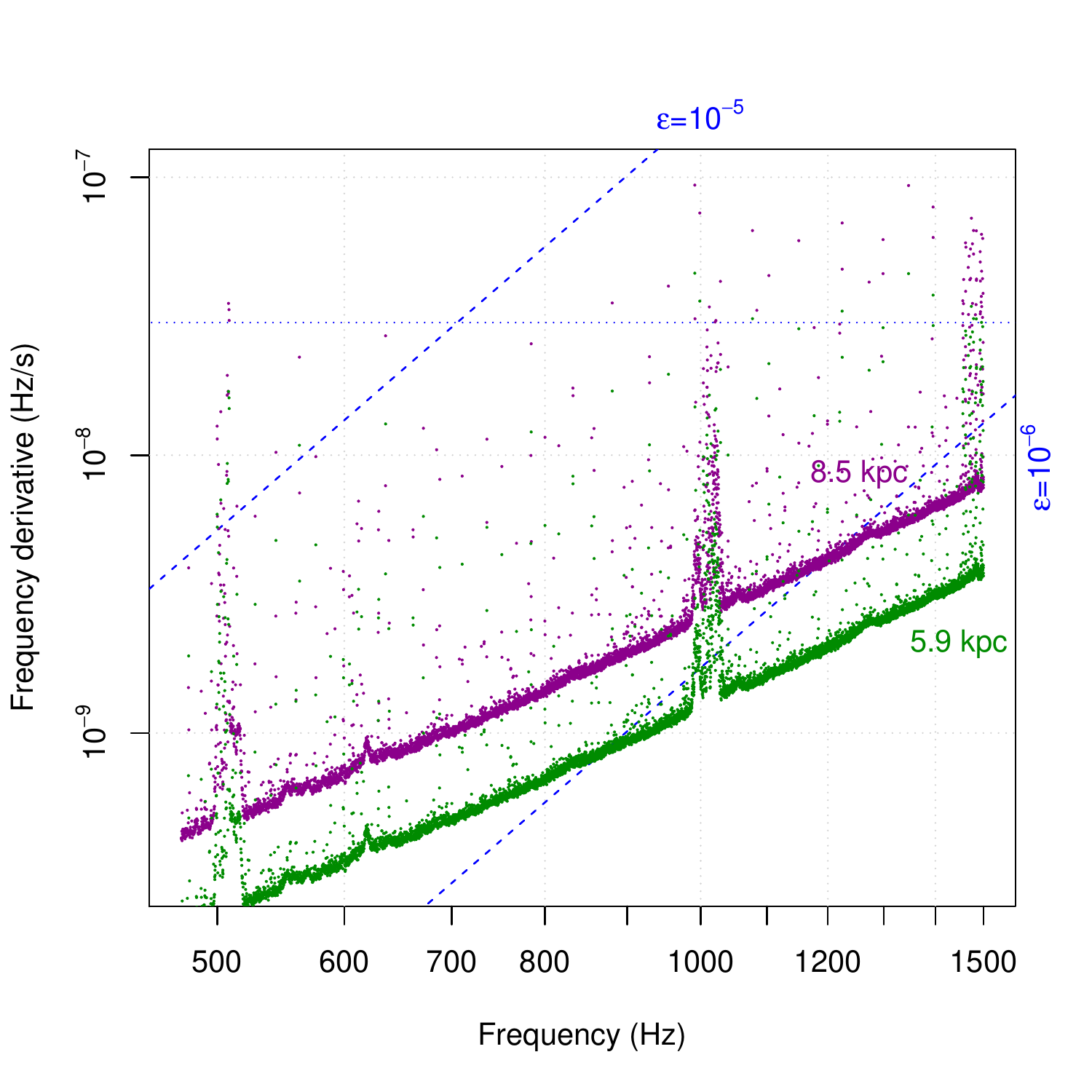}
\caption[Spindown range]{
\label{fig:spindown_range}
Range of the search for neutron stars
spinning down solely due to gravitational radiation.  This is a
superposition of two contour plots.  The green (bottom) and purple (top) solid markers are contours of the maximum distance at which a neutron
star could be detected as a function of gravitational-wave frequency
$f$ and its derivative $\dot{f}$.  The dashed lines 
are contours of the corresponding ellipticity
$\epsilon(f,\dot{f})$. The fine dotted line marks the maximum spindown searched. Together these quantities tell us the
maximum range of the search in terms of various populations (see text
for details) (color online). 5.9 kpc is the distance associated to many ATNF catalogue pulsars in the direction of Terzan 5.
}
\end{figure}

Figure \ref{fig:spindown_range} shows the astrophysical reach of the search, i.e. the maximum distance at which this search could have detected a signal of a given frequency and spin-down, under the assumption that all the lost rotational energy is emitted in gravitational waves. The search presented here is sensitive to an optimally oriented neutron star at the galactic center (circularly polarized signal) with ellipticity of $10^{-6}$ and emitting gravitational waves at a frequency of $1200$\,Hz. In Terzan 5 a signal at $1200$\,Hz from an optimally oriented source having ellipticity of $\leq \sci{7}{-7}$ could have been detected. 

The search presented is the most sensitive to date, aimed at this interesting region of our galaxy. This is reflected in the sensitivity depth of the search which is defined as the ratio of the upper limit value and the noise floor at nearby frequencies ${{\cal{D}}(f)}:=\sqrt{S_h(f)\over h^{UL}_0}$ \cite{SensDepthDef}. Following \cite{Dreissigacker:2018afk}, we estimate the noise taking the harmonic mean across the different detectors and obtain the following values of the sensitivity depth across the entire frequency range searched:
\begin{equation}
\begin{cases}
{{\cal{D}}_{\texttt{circ-pol}}(f)} &= 116 ~~[\sqrt{{\textrm{Hz}}}]^{-1/2}\\
{{\cal{D}}_{\texttt{worst-pol}}(f)} &= 42~~[\sqrt{{\textrm{Hz}}}]^{-1/2}.
\end{cases}
\label{eq:FU1region}
\end{equation}
The radiometer search  \cite{LVCRadiometerO1} targeting the galactic center is 4 times less sensitive than our most conservative upper limit (the worst case one), achieving, on the same data, a sensitivity depth smaller than 10.
This search covers a larger spindown range than any previously published all-sky search, hence probing younger sources from our search area. Furthermore even our worst-case upper limits are more constraining than any all-sky search result to date,  including the state of the art paper \cite{O2AllSkyLVC} that uses the more sensitive and longer duration data set from the O2 run.

\section{Acknowledgments}

The simulations and data analysis were performed on Atlas cluster at AEI Hannover, for which we thank Bruce Allen. We thank Carsten Aulbert and Henning Fehrmann for their support.

We thank Teviet Creighton and Keith Riles for many helpful comments and encouragement over the years.

The authors thank the LIGO Scientific Collaboration for access to the data and gratefully acknowledge the support of the United States National Science Foundation (NSF) for the construction and operation of the LIGO Laboratory and Advanced LIGO as well as the Science and Technology Facilities Council (STFC) of the United Kingdom, and the Max-Planck-Society (MPS) for support of the construction of Advanced LIGO. Additional support for Advanced LIGO was provided by the Australian Research Council.

This research has made use of data, software and/or web tools obtained from the LIGO Open Science Center (\url{https://losc.ligo.org}), a service of LIGO Laboratory, the LIGO Scientific Collaboration and the Virgo Collaboration. LIGO is funded by the U.S. National Science Foundation. Virgo is funded by the French Centre National de Recherche Scientifique (CNRS), the Italian Istituto Nazionale della Fisica Nucleare (INFN) and the Dutch Nikhef, with contributions by Polish and Hungarian institutes.

\appendix
\section{Outlier tables}
\label{sec:outliers}

Outliers passing all stages of automated followup from 475-1500\,Hz band are separated into five tables.
Table \ref{tab:PowerFluxOutliersViolin} shows outliers inside the contaminated regions 495-520\,Hz and 990-1033\,Hz. The rest of the outliers is split into four regions
475-900\,Hz, 900-1200\,Hz, 1200-1400\,Hz and 1400-1500\,Hz (Tables \ref{tab:PowerFluxOutliers1}, \ref{tab:PowerFluxOutliers2}, \ref{tab:PowerFluxOutliers3}, \ref{tab:PowerFluxOutliers4}).

\begin{table*}[htbp]
\begin{center}
\scriptsize
\begin{tabular}{D{.}{.}{2}D{.}{.}{2}cD{.}{.}{5}D{.}{.}{4}D{.}{.}{4}D{.}{.}{4}l}\hline
\multicolumn{1}{c}{Idx} & \multicolumn{1}{c}{SNR}  & \multicolumn{1}{c}{$\log_{10}\left({\textrm{GFA}}\right)$} &  \multicolumn{1}{c}{Frequency} & \multicolumn{1}{c}{Spindown} &  \multicolumn{1}{c}{$\RAJ$}  & \multicolumn{1}{c}{$\DECJ$} & Description \\
\multicolumn{1}{c}{}	&  \multicolumn{1}{c}{}	&  \multicolumn{1}{c}{}	& \multicolumn{1}{c}{Hz}	&  \multicolumn{1}{c}{nHz/s} & \multicolumn{1}{c}{degrees} & \multicolumn{1}{c}{degrees} & \\
\hline \hline
6 & 23.9 & -17.5 & 612.48610 & -9.197 & 267.037 & -29.754 & Broad large line in L1 at 612.45 Hz \\
8 & 23.5 & -16.6 & 736.09475 & -22.991 & 267.808 & -28.287 & Sharp bin-centered lines at 736 Hz (H1) and 736.1 (L1) \\
13 & 21.2 & -11.7 & 736.09791 & -21.997 & 266.439 & -29.165 & Sharp bin-centered lines at 736 Hz (H1) and 736.1 (L1) \\
17 & 20.3 & -9.6 & 684.96515 & -17.607 & 265.576 & -22.464 & Sharp bin-centered line in L1 at 684.9 Hz \\
20 & 19.2 & -7.5 & 662.18356 & -24.621 & 263.548 & -25.681 & Strong bin-centered line in L1 at 662.20 Hz \\
24 & 17.9 & -4.8 & 710.54465 & -21.826 & 264.817 & -27.547 & Strong bin-centered line in L1 at 710.50 Hz \\
32 & 17.6 & -4.0 & 599.19367 & -15.234 & 266.039 & -28.665 & Large broad lines in H1 near 599.14 Hz and 599.16 Hz \\
34 & 17.5 & -3.8 & 761.75580 & -13.218 & 265.953 & -25.889 & Strong bin-centered line in L1 at 761.70 Hz (FU0) \\
43 & 16.8 & -2.4 & 707.65162 & -10.129 & 266.799 & -24.227 & Strong bin-centered line in L1 at 707.6 Hz \\
44 & 16.6 & -2.1 & 575.23174 & -5.519 & 266.534 & -29.341 & Hardware injected pulsar 2 \\
45 & 16.6 & -2.0 & 898.86970 & -0.491 & 269.397 & -28.914 & Large broad lines in H1 \\
47 & 16.5 & -1.9 & 898.84667 & -15.157 & 266.790 & -26.909 & Large broad lines in H1 \\
70 & 15.2 & 0.8 & 659.35418 & -12.386 & 269.229 & -26.124 & Strong bin-centered line in L1 at 659.3 Hz \\
72 & 15.1 & 0.9 & 629.86431 & -29.241 & 267.817 & -25.916 & Large broad lines in L1 \\
75 & 15.0 & 1.1 & 787.35687 & -8.004 & 265.084 & -29.806 & Sharp bin-centered line in L1 at 787.3 Hz \\
80 & 14.8 & 1.4 & 660.51361 & -20.472 & 268.538 & -22.492 & \\
84 & 14.8 & 1.5 & 829.85946 & -8.609 & 264.856 & -25.815 & Strong bin-centered line in L1 at 829.8 Hz \\
85 & 14.8 & 1.5 & 520.84815 & -3.204 & 267.198 & -29.179 & Large broad line in H1 near 520.82 Hz \\
89 & 14.7 & 1.7 & 520.84814 & -3.201 & 267.202 & -29.359 & Large broad line in H1 near 520.82 Hz \\
99 & 14.5 & 2.1 & 763.94306 & -20.331 & 262.960 & -26.415 & Hardware injected pulsar 9 \\
102 & 14.5 & 2.1 & 873.26713 & -29.286 & 264.627 & -28.210 & (FU0)\\
103 & 14.5 & 2.1 & 606.63606 & -27.504 & 263.085 & -27.581 & Large broad line in H1 at 606.67 Hz \\
104 & 14.4 & 2.3 & 730.35349 & -8.153 & 266.654 & -26.898 & Sharp bin-centered line in L1 at 730.3 Hz \\
119 & 14.2 & 2.7 & 787.35542 & -8.931 & 264.431 & -28.451 & Sharp bin-centered line in L1 at 787.3 Hz \\
120 & 14.1 & 2.8 & 608.06595 & -19.469 & 268.412 & -28.434 & Sharp bin-centered line in H1 at 608 Hz \\
121 & 14.1 & 2.8 & 899.25624 & -27.714 & 263.423 & -28.511 & Strong broad lines in H1 \\
135 & 14.0 & 3.1 & 599.49600 & -17.127 & 267.287 & -28.898 & Strong broad lines in H1\\
137 & 14.0 & 3.1 & 771.05117 & -8.789 & 265.082 & -22.538 & (FU1) \\
139 & 13.9 & 3.1 & 587.37228 & -0.374 & 265.864 & -29.281 & (FU1) \\
145 & 13.9 & 3.2 & 864.06026 & -9.286 & 262.769 & -25.668 & Sharp bin-centered line in H1 at 864 Hz \\
146 & 13.9 & 3.3 & 575.23399 & -5.628 & 268.609 & -27.074 & Hardware injected pulsar 2 \\
149 & 13.9 & 3.3 & 764.65686 & -8.411 & 267.769 & -27.209 & Sharp bin-centered line in L1 at 764.6 Hz \\
151 & 13.9 & 3.3 & 817.31637 & -23.376 & 267.491 & -28.124 & (FU1) \\
156 & 13.8 & 3.4 & 773.67502 & -4.093 & 265.308 & -23.202 & \\
161 & 13.8 & 3.5 & 618.06167 & -13.849 & 268.606 & -25.285 & Slope in H1 spectrum \\
166 & 13.8 & 3.5 & 738.02255 & -0.388 & 263.204 & -27.112 & (FU0)\\
169 & 13.7 & 3.6 & 629.86433 & -29.236 & 267.845 & -26.116 & Large broad line in L1 \\
170 & 13.7 & 3.6 & 769.35342 & -26.944 & 267.449 & -22.402 & (FU1) \\
172 & 13.7 & 3.6 & 686.75565 & -11.731 & 264.065 & -29.473 & (FU0)\\
176 & 13.7 & 3.7 & 764.65687 & -8.411 & 267.773 & -27.188 & Sharp bin-centered line in L1 at 764.6 Hz \\
180 & 13.6 & 3.7 & 683.40267 & -12.421 & 265.532 & -28.007 & Slight slope in L1 (FU0)\\
186 & 13.6 & 3.8 & 799.26703 & -5.724 & 266.255 & -28.143 & (FU0)\\
190 & 13.6 & 3.8 & 824.59028 & -2.196 & 268.212 & -27.435 & (FU0)\\
192 & 13.6 & 3.8 & 645.94631 & -15.366 & 266.741 & -26.423 & \\
202 & 13.5 & 3.9 & 727.32568 & -17.438 & 267.008 & -27.029 & (FU1) \\
203 & 13.5 & 3.9 & 539.85863 & -8.429 & 267.373 & -29.396 & Near 60 Hz line \\
215 & 13.5 & 4.1 & 851.68971 & -18.266 & 266.383 & -22.291 & (FU1) \\
219 & 13.4 & 4.1 & 489.11959 & -5.482 & 265.242 & -25.347 & Nearby lines \\
227 & 13.4 & 4.2 & 694.42637 & -26.409 & 267.828 & -27.898 & (FU0)\\
229 & 13.4 & 4.2 & 581.71075 & -9.491 & 263.276 & -28.388 & (FU1) \\
232 & 13.4 & 4.3 & 713.46388 & -2.209 & 266.696 & -26.761 & Strong bin-centered line in L1 at 713.4 Hz \\
243 & 13.3 & 4.3 & 575.25658 & -18.581 & 265.147 & -29.940 & Hardware injected pulsar 2 \\
244 & 13.3 & 4.3 & 583.35870 & -8.394 & 269.473 & -29.052 & Strong broad line in H1 at 583.317 (FU0)\\
248 & 13.3 & 4.4 & 763.95114 & -21.211 & 266.312 & -24.681 & Hardware injected pulsar 9 \\
249 & 13.3 & 4.4 & 680.27362 & -14.639 & 263.404 & -25.497 & (FU0)\\
251 & 13.3 & 4.4 & 608.00120 & -16.846 & 263.648 & -28.505 & Sharp bin-centered line in H1 at 608 Hz \\
254 & 13.3 & 4.4 & 770.00605 & -19.542 & 264.245 & -27.578 & (FU1) \\
260 & 13.3 & 4.4 & 772.83474 & -27.344 & 267.338 & -26.027 & Sharp bin-centered line in L1 at 772.8 Hz\\
261 & 13.3 & 4.4 & 809.98476 & -27.979 & 263.634 & -25.694 & Sharp bin-centered line in L1 at 810 Hz \\
264 & 13.3 & 4.4 & 694.42638 & -26.413 & 267.818 & -27.711 & (FU0)\\
270 & 13.3 & 4.5 & 878.16583 & -14.756 & 267.454 & -26.247 & Strong bin-centered line in L1 at 878.1 Hz\\
282 & 13.2 & 4.5 & 547.67510 & -9.729 & 263.964 & -27.205 & (FU0)\\
287 & 13.2 & 4.6 & 829.86026 & -8.384 & 265.709 & -27.241 & Strong bin-centered line in L1 at 829.8 Hz \\
290 & 13.2 & 4.6 & 792.70263 & -16.328 & 266.224 & -29.232 & (FU0)\\
291 & 13.2 & 4.6 & 655.43480 & -2.403 & 263.896 & -25.645 & (FU0)\\
297 & 13.2 & 4.6 & 848.06291 & -16.696 & 264.046 & -26.114 & Strong bin-centered line in H1 at 848 Hz \\
299 & 13.2 & 4.6 & 725.52768 & -5.731 & 263.854 & -25.707 & (FU0)\\
303 & 13.2 & 4.6 & 782.79214 & -20.134 & 265.043 & -26.320 & (FU1) \\
309 & 13.1 & 4.7 & 599.20398 & -18.747 & 269.644 & -26.805 & Big broad lines in H1 near 599.14 and 599.16 Hz \\
310 & 13.1 & 4.7 & 698.22224 & -22.943 & 263.921 & -25.129 & (FU0)\\
311 & 13.1 & 4.7 & 763.95114 & -21.208 & 266.313 & -24.999 & Hardware injected pulsar 9 \\
313 & 13.1 & 4.7 & 527.12774 & -7.311 & 266.075 & -25.556 & (FU0)\\
316 & 13.1 & 4.8 & 753.97528 & 0.849 & 266.035 & -25.250 & (FU0)\\
317 & 13.1 & 4.8 & 844.15841 & -9.201 & 262.865 & -24.558 & (FU1) \\
319 & 13.1 & 4.8 & 799.63851 & -3.359 & 265.959 & -29.050 & (FU1) \\
323 & 13.1 & 4.8 & 718.02858 & -22.968 & 266.134 & -27.019 & (FU1) \\
325 & 13.1 & 4.8 & 527.64148 & -23.659 & 268.915 & -24.650 & (FU0)\\
335 & 13.0 & 4.9 & 621.85099 & -11.539 & 266.052 & -27.996 & Sloping H1 spectrum \\
337 & 13.0 & 4.9 & 676.37758 & -0.769 & 268.788 & -24.882 & (FU0)\\
340 & 13.0 & 4.9 & 678.39254 & -3.871 & 263.974 & -28.642 & \\

\hline
\end{tabular}
\caption[Outliers that passed the automated detection pipeline]{Outliers below 900\,Hz that passed the automated detection pipeline excluding regions heavily contaminated with violin modes. Outliers marked with ``line'' had strong narrowband disturbances identified near the outlier location. We have marked outliers not consistent with the target signals at one of the semi-coherent \Fstat ~ follow-ups with ``(FU0/1/2)", depending on the stage at which they did not pass the detection thresholds. 
Frequencies are converted to epoch GPS $1130529362$.}
\label{tab:PowerFluxOutliers1}
\end{center}
\end{table*}

\begin{table*}[htbp]
\begin{center}
\scriptsize
\begin{tabular}{D{.}{.}{2}D{.}{.}{2}cD{.}{.}{5}D{.}{.}{4}D{.}{.}{4}D{.}{.}{4}l}\hline
\multicolumn{1}{c}{Idx} & \multicolumn{1}{c}{SNR}  & \multicolumn{1}{c}{$\log_{10}\left({\textrm{GFA}}\right)$} &  \multicolumn{1}{c}{Frequency} & \multicolumn{1}{c}{Spindown} &  \multicolumn{1}{c}{$\RAJ$}  & \multicolumn{1}{c}{$\DECJ$} & Description \\
\multicolumn{1}{c}{}	&  \multicolumn{1}{c}{}	&  \multicolumn{1}{c}{}	& \multicolumn{1}{c}{Hz}	&  \multicolumn{1}{c}{nHz/s} & \multicolumn{1}{c}{degrees} & \multicolumn{1}{c}{degrees} & \\
\hline \hline
5 & 25.2 & -20.4 & 1176.69799 & -26.024 & 264.288 & -23.567 & Strong bin-centered line in L1 at 1176.6 Hz \\ 
16 & 20.8 & -10.8 & 955.00851 & -26.744 & 269.405 & -24.278 & Sharp line in L1 \\
19 & 19.3 & -7.6 & 910.17153 & -8.556 & 264.028 & -24.077 & Large broad line in H1 at 910.1 Hz \\
33 & 17.5 & -4.0 & 1176.58614 & -3.254 & 263.501 & -25.977 & Strong bin-centered line in L1 at 1176.6 Hz \\
39 & 17.0 & -2.9 & 1120.09915 & -25.758 & 264.594 & -27.448 & Strong bin-centered line in H1 at 1120 Hz \\
40 & 16.9 & -2.6 & 910.18376 & -22.563 & 268.269 & -25.989 & Large broad line in H1 at 910.1 Hz \\
58 & 15.5 & 0.1 & 1173.80211 & -28.189 & 264.045 & -27.467 & Strong bin-centered line in L1 at 1173.7 Hz \\
59 & 15.5 & 0.1 & 1128.38343 & -12.747 & 267.927 & -27.526 & Strong bin-centered line in L1 at 1128.3 Hz \\
63 & 15.3 & 0.4 & 906.63379 & -17.648 & 264.077 & -23.762 & Strong bin-centered line in L1 at 906.6 Hz \\
68 & 15.2 & 0.6 & 1105.15733 & -26.774 & 263.713 & -26.330 &  (FU1 w. O2 data)\\
76 & 15.0 & 1.2 & 1128.39396 & -25.644 & 263.670 & -27.537 & Strong bin-centered line in L1 1128.3 Hz \\
77 & 14.9 & 1.3 & 946.92321 & -20.521 & 266.068 & -27.668 & (FU0)\\
81 & 14.8 & 1.4 & 977.67244 & -18.241 & 264.215 & -26.981 & Strong bin-centered line in L1 at 977.6 Hz \\
86 & 14.7 & 1.6 & 1130.32268 & -1.751 & 267.651 & -29.016 & (FU0)\\
101 & 14.5 & 2.1 & 983.46951 & -11.611 & 264.204 & -23.932 & Strong bin-centered line in L1 at 983.4 Hz \\
106 & 14.4 & 2.3 & 976.01956 & -26.851 & 265.826 & -29.672 & Line in H1 at 976 Hz \\
113 & 14.3 & 2.5 & 957.88553 & -2.921 & 266.766 & -24.376 & Sharp bin-centered line in L1 at 957.8 Hz \\
114 & 14.3 & 2.5 & 932.28876 & -0.777 & 265.303 & -25.123 & Sharp bin-centered line in L1 at 932.2 Hz \\
116 & 14.2 & 2.6 & 1083.85197 & -19.421 & 266.328 & -24.585 & (FU0)\\
118 & 14.2 & 2.7 & 1117.93710 & -29.371 & 266.174 & -24.014 & \\
125 & 14.1 & 2.9 & 1192.54799 & 0.039 & 267.676 & -26.744 & (FU0)\\
126 & 14.1 & 2.9 & 1144.81799 & -12.261 & 265.508 & -28.590 & (FU0)\\
130 & 14.0 & 3.0 & 1146.01080 & -14.459 & 264.505 & -26.227 & (FU1) \\
133 & 14.0 & 3.1 & 1056.49144 & -6.807 & 263.080 & -25.072 & (FU1) \\
143 & 13.9 & 3.2 & 916.79125 & -21.067 & 266.033 & -24.889 & (FU0)\\
147 & 13.9 & 3.3 & 1055.06400 & -21.219 & 267.870 & -25.985 & (FU0)\\
148 & 13.9 & 3.3 & 1148.12864 & -9.921 & 268.460 & -28.864 & Strong bin-centered line in L1 at 1148.1 Hz \\
158 & 13.8 & 3.4 & 1193.00546 & -15.631 & 266.493 & -25.264 & (FU0)\\
159 & 13.8 & 3.5 & 911.76958 & -9.399 & 266.039 & -25.180 & (FU0)\\
160 & 13.8 & 3.5 & 1130.73154 & -3.169 & 263.000 & -26.335 & (FU0)\\
162 & 13.8 & 3.5 & 953.40039 & -10.813 & 266.270 & -27.596 & \\
174 & 13.7 & 3.6 & 1087.75530 & -28.077 & 269.614 & -26.525 & (FU2)  \\
181 & 13.6 & 3.7 & 969.52238 & -8.313 & 265.371 & -29.459 & (FU0) \\
206 & 13.5 & 4.0 & 1159.91542 & -24.221 & 264.096 & -25.279 & (FU0)\\
208 & 13.5 & 4.0 & 1142.86654 & -6.054 & 263.246 & -24.169 & (FU0)\\
210 & 13.5 & 4.0 & 934.78261 & -2.273 & 268.362 & -25.152 & (FU0)\\
211 & 13.5 & 4.0 & 1080.01473 & -27.759 & 266.170 & -28.283 & Strong coincident bin-centered lines in H1 and L1 at 1080 Hz \\
213 & 13.5 & 4.1 & 1127.80227 & -19.557 & 267.194 & -23.981 & (FU1) \\
223 & 13.4 & 4.2 & 970.66243 & -29.248 & 266.863 & -25.559 & (FU0)\\
228 & 13.4 & 4.2 & 931.29979 & -10.836 & 266.227 & -29.987 & \\
233 & 13.4 & 4.3 & 1151.53614 & -8.588 & 265.597 & -23.441 & \\
237 & 13.3 & 4.3 & 1145.63773 & -5.498 & 268.460 & -29.377 & (FU0)\\
242 & 13.3 & 4.3 & 983.47659 & -7.236 & 265.003 & -27.225 & Strong bin-centered line in L1 at 983.4 Hz \\
245 & 13.3 & 4.3 & 1197.77064 & -27.963 & 269.015 & -27.561 & \\
253 & 13.3 & 4.4 & 903.29002 & -15.369 & 266.450 & -25.044 & (FU0)\\
258 & 13.3 & 4.4 & 938.88434 & -9.556 & 264.187 & -28.599 & (FU0)\\
265 & 13.3 & 4.4 & 953.40039 & -10.816 & 266.269 & -27.442 & \\
266 & 13.3 & 4.5 & 906.90104 & -20.814 & 263.359 & -26.506 & Broad line in H1 near 906.82 Hz \\
267 & 13.3 & 4.5 & 1069.13874 & -2.926 & 263.432 & -27.102 & \\
272 & 13.2 & 4.5 & 1033.96710 & -20.787 & 262.779 & -25.972 & (FU0)\\
273 & 13.2 & 4.5 & 1039.41470 & 0.882 & 267.926 & -24.697 & \\
276 & 13.2 & 4.5 & 1121.71865 & -15.156 & 263.124 & -27.755 & (FU0)\\
279 & 13.2 & 4.5 & 1102.25911 & -26.261 & 266.982 & -27.473 & (FU1) \\
284 & 13.2 & 4.5 & 1055.96111 & -20.166 & 266.229 & -28.653 & (FU0)\\
285 & 13.2 & 4.5 & 1081.91076 & -29.388 & 264.805 & -29.647 & (FU0)\\
295 & 13.2 & 4.6 & 1143.15676 & -5.409 & 269.119 & -26.704 & (FU1) \\
300 & 13.2 & 4.6 & 951.01213 & -4.077 & 266.334 & -28.538 & (FU0)\\
305 & 13.2 & 4.7 & 1070.50637 & -4.409 & 264.544 & -27.297 & \\
307 & 13.1 & 4.7 & 1123.01894 & -10.841 & 268.163 & -24.666 & (FU0)\\
314 & 13.1 & 4.7 & 945.03386 & -24.222 & 267.786 & -26.393 & Bump in L1 \\
318 & 13.1 & 4.8 & 989.77830 & -4.182 & 269.861 & -25.516 & Disturbed H1 spectrum \\
330 & 13.1 & 4.8 & 1176.33662 & -22.274 & 264.698 & -27.271 & (FU1) \\
336 & 13.0 & 4.9 & 985.14327 & -23.229 & 265.622 & -26.558 & Many strong nearby lines in H1 \\
338 & 13.0 & 4.9 & 1090.81873 & -0.502 & 264.444 & -29.353 & (FU0)\\
339 & 13.0 & 4.9 & 1196.00279 & -2.451 & 267.251 & -24.656 & (FU0)\\
342 & 13.0 & 4.9 & 1037.60585 & -26.743 & 263.801 & -28.995 & Strong bin-centered line in L1 at 1037.5 Hz \\
350 & 13.0 & 4.9 & 1197.48407 & -23.759 & 269.911 & -25.523 & \\

\hline
\end{tabular}
\caption[Outliers that passed detection pipeline]{Outliers in frequency range 900-1200\,Hz that passed the detection pipeline excluding regions heavily contaminated with violin modes. Outliers marked with ``line'' had strong narrowband disturbances identified near the outlier location. We have marked outliers not consistent with the target signals at one of the semi-coherent \Fstat ~ follow-ups with ``(FU0/1/2)", depending on the stage at which they did not pass the detection thresholds.
Frequencies are converted to epoch GPS $1130529362$.}
\label{tab:PowerFluxOutliers2}
\end{center}
\end{table*}

\begin{table*}[htbp]
\begin{center}
\scriptsize
\begin{tabular}{D{.}{.}{2}D{.}{.}{2}cD{.}{.}{5}D{.}{.}{4}D{.}{.}{4}D{.}{.}{4}l}\hline
\multicolumn{1}{c}{Idx} & \multicolumn{1}{c}{SNR}  & \multicolumn{1}{c}{$\log_{10}\left({\textrm{GFA}}\right)$} &  \multicolumn{1}{c}{Frequency} & \multicolumn{1}{c}{Spindown} &  \multicolumn{1}{c}{$\RAJ$}  & \multicolumn{1}{c}{$\DECJ$} & Description \\
\multicolumn{1}{c}{}	&  \multicolumn{1}{c}{}	&  \multicolumn{1}{c}{}	& \multicolumn{1}{c}{Hz}	&  \multicolumn{1}{c}{nHz/s} & \multicolumn{1}{c}{degrees} & \multicolumn{1}{c}{degrees} & \\
\hline \hline
1 & 30.5 & -32.4 & 1220.62344 & -16.068 & 265.994 & -24.436 & Induced by hardware injection 7 (FU0) \\
37 & 17.0 & -2.9 & 1360.09284 & -16.252 & 262.922 & -27.103 & Strong bin-centered line in H1 at 1360 Hz \\
42 & 16.8 & -2.5 & 1276.22672 & -0.304 & 268.176 & -23.187 & Strong bin-centered line in L1 at 1276.1 Hz \\
55 & 15.8 & -0.4 & 1202.29927 & -6.043 & 264.510 & -24.894 & Strong bin-centered line in L1 at 1202.2 Hz \\
56 & 15.7 & -0.2 & 1376.12253 & -4.253 & 269.645 & -26.036 & Strong bin-centered line in H1 at 1376 Hz \\
67 & 15.3 & 0.6 & 1280.12932 & 0.519 & 268.744 & -23.550 & Strong bin-centered line in H1 at 1280 Hz \\
69 & 15.2 & 0.8 & 1270.39556 & -12.976 & 268.516 & -24.920 & Strong bin-centered line in L1 at 1270.3 Hz \\
78 & 14.9 & 1.3 & 1328.13275 & 0.159 & 269.294 & -26.578 & Strong bin-centered line in H1 at 1328 Hz \\
82 & 14.8 & 1.5 & 1202.29985 & -5.886 & 264.920 & -25.698 & Strong bin-centered line in L1 at 1202.2 Hz \\
91 & 14.7 & 1.8 & 1376.09811 & -20.913 & 267.292 & -26.665 & Strong bin-centered line in H1 at 1376 Hz (FU0)\\
93 & 14.6 & 1.8 & 1355.19792 & -22.717 & 265.274 & -27.347 & (FU1) \\
96 & 14.6 & 2.0 & 1303.93001 & -13.662 & 268.017 & -29.082 & (FU0)\\
107 & 14.3 & 2.4 & 1321.58965 & -19.634 & 263.445 & -25.948 & Strong bin-centered line in L1 at 1321.5 Hz \\
111 & 14.3 & 2.5 & 1352.11873 & -4.071 & 266.570 & -28.811 & Bin-centered line in H1 at 1352 Hz \\
112 & 14.3 & 2.5 & 1254.44064 & -15.222 & 265.991 & -25.692 & (FU1) \\
115 & 14.2 & 2.6 & 1301.62221 & -15.454 & 265.848 & -28.031 & \\
117 & 14.2 & 2.6 & 1202.30356 & -26.059 & 267.238 & -23.480 & Strong bin-centered line in L1 at 1202.2 Hz \\
123 & 14.1 & 2.9 & 1270.39317 & -17.622 & 268.335 & -25.117 & Strong bin-centered line in L1 at 1270.3 Hz \\
127 & 14.0 & 3.0 & 1386.49465 & -13.231 & 267.856 & -23.177 & (FU0)\\
128 & 14.0 & 3.0 & 1380.26131 & -19.417 & 263.195 & -24.722 & (FU0)\\
129 & 14.0 & 3.0 & 1264.07644 & -19.604 & 265.130 & -29.366 & Line in at 1264 Hz in H1 \\
138 & 13.9 & 3.1 & 1249.43252 & -2.559 & 265.309 & -23.949 &   (FU0)\\
140 & 13.9 & 3.2 & 1373.86300 & -7.084 & 263.597 & -27.357 & (FU0)\\
141 & 13.9 & 3.2 & 1205.72493 & -26.927 & 264.610 & -28.241 & (FU0)\\
142 & 13.9 & 3.2 & 1271.07364 & -17.608 & 266.148 & -28.676 & (FU0)\\
144 & 13.9 & 3.2 & 1366.55954 & -0.596 & 262.870 & -26.589 & (FU0)\\
150 & 13.9 & 3.3 & 1331.30890 & -21.827 & 263.364 & -28.484 & (FU1) \\
152 & 13.8 & 3.4 & 1264.10309 & -6.094 & 262.596 & -26.501 & Strong bin-centered line in H1 at 1264 Hz \\
163 & 13.8 & 3.5 & 1315.28928 & -21.894 & 270.072 & -25.232 & (FU0)\\
165 & 13.8 & 3.5 & 1380.26131 & -19.417 & 263.200 & -24.633 & (FU0)\\
171 & 13.7 & 3.6 & 1269.91895 & -21.536 & 267.428 & -28.206 & (FU0)\\
173 & 13.7 & 3.6 & 1276.99016 & -0.241 & 265.645 & -29.501 & (FU1) \\
175 & 13.7 & 3.7 & 1332.83814 & -16.049 & 265.781 & -27.931 & (FU0)\\
177 & 13.7 & 3.7 & 1372.18144 & -19.329 & 264.461 & -27.028 & (FU1) \\
179 & 13.7 & 3.7 & 1267.04168 & -8.514 & 268.806 & -23.928 & (FU0)\\
182 & 13.6 & 3.7 & 1232.09595 & -8.929 & 266.351 & -26.068 & Strong bin-centered line in H1 at 1232 Hz \\
188 & 13.6 & 3.8 & 1254.44064 & -15.229 & 265.990 & -25.311 & (FU1) \\
189 & 13.6 & 3.8 & 1394.57034 & -8.916 & 269.267 & -23.752 & (FU1) \\
191 & 13.6 & 3.8 & 1270.38626 & -19.122 & 263.603 & -25.026 & Strong bin-centered line in L1 at 1270.3 Hz \\
194 & 13.6 & 3.8 & 1318.61537 & -2.694 & 263.896 & -27.405 & Line in L1 at 1318.6 Hz ??? \\
195 & 13.6 & 3.9 & 1262.65832 & -14.294 & 266.151 & -27.129 & (FU0) \\
199 & 13.5 & 3.9 & 1212.89760 & -9.841 & 262.769 & -25.319 & (FU1) \\
207 & 13.5 & 4.0 & 1202.29841 & -27.188 & 263.136 & -26.481 & Strong bin-centered line in L1 at 1202.2 Hz \\
214 & 13.5 & 4.1 & 1232.09153 & -18.288 & 268.914 & -24.416 & Strong bin-centered line in H1 at 1232 Hz \\
218 & 13.5 & 4.1 & 1270.38624 & -19.118 & 263.585 & -25.347 & Strong bin-centered line in L1 at 1270.3 Hz \\
220 & 13.4 & 4.1 & 1313.36532 & 0.507 & 266.019 & -23.381 & (FU0)\\
236 & 13.3 & 4.3 & 1242.80188 & 0.099 & 263.747 & -28.690 & (FU0)\\
238 & 13.3 & 4.3 & 1210.06445 & -0.084 & 264.836 & -26.046 & \\
241 & 13.3 & 4.3 & 1317.50608 & -14.391 & 267.541 & -28.868 & (FU0)\\
246 & 13.3 & 4.3 & 1280.91794 & -20.272 & 265.138 & -29.732 & (FU1) \\
247 & 13.3 & 4.4 & 1222.09162 & -9.369 & 268.642 & -26.238 & Strong bin-centered line in L1 at 1222 Hz\\
255 & 13.3 & 4.4 & 1253.01107 & -5.959 & 265.096 & -25.776 & \\
257 & 13.3 & 4.4 & 1323.77888 & 0.764 & 268.106 & -29.354 & (FU0)\\
262 & 13.3 & 4.4 & 1253.01107 & -5.964 & 265.098 & -25.512 & (FU0)\\
263 & 13.3 & 4.4 & 1261.98690 & -6.648 & 265.929 & -24.334 & (FU0)\\
269 & 13.3 & 4.5 & 1329.92157 & -2.304 & 264.893 & -25.113 & (FU0)\\
271 & 13.2 & 4.5 & 1371.33276 & -6.367 & 267.733 & -27.949 & (FU0)\\
274 & 13.2 & 4.5 & 1299.81900 & -3.076 & 265.372 & -26.913 & (FU0)\\
278 & 13.2 & 4.5 & 1289.14048 & -18.428 & 265.039 & -27.429 & (FU0)\\
283 & 13.2 & 4.5 & 1292.03374 & -18.191 & 263.987 & -24.625 & (FU0)\\
286 & 13.2 & 4.6 & 1298.51021 & -3.954 & 269.268 & -26.330 & (FU0)\\
288 & 13.2 & 4.6 & 1321.82947 & -26.962 & 267.202 & -26.389 & \\
289 & 13.2 & 4.6 & 1399.83041 & -29.214 & 264.793 & -29.091 & \\
293 & 13.2 & 4.6 & 1247.68962 & -21.128 & 265.493 & -23.383 & Strong bin-centered line in L1 at 1247.6 Hz (FU0)\\
294 & 13.2 & 4.6 & 1225.38754 & -28.327 & 264.042 & -25.595 & (FU0)\\
298 & 13.2 & 4.6 & 1359.45910 & -2.732 & 266.847 & -26.802 & (FU0)\\
308 & 13.1 & 4.7 & 1282.33163 & -13.246 & 263.533 & -26.441 & (FU0)\\
322 & 13.1 & 4.8 & 1225.72267 & -17.528 & 263.821 & -25.398 & (FU0)\\
327 & 13.1 & 4.8 & 1249.95109 & -23.121 & 266.789 & -27.051 & (FU1) \\
331 & 13.1 & 4.8 & 1297.00050 & -1.628 & 267.856 & -26.736 & (FU0)\\
332 & 13.1 & 4.8 & 1215.84097 & -17.271 & 269.914 & -27.022 & (FU0)\\
333 & 13.0 & 4.9 & 1295.23973 & -11.871 & 266.940 & -28.091 & (FU0)\\
334 & 13.0 & 4.9 & 1306.38413 & -18.809 & 265.273 & -22.916 &(FU0) \\
341 & 13.0 & 4.9 & 1345.96737 & -26.084 & 263.623 & -26.514 & (FU1) \\
343 & 13.0 & 4.9 & 1216.27144 & -10.614 & 267.855 & -25.711 & (FU0)\\
344 & 13.0 & 4.9 & 1259.45564 & -9.417 & 267.383 & -28.047 & (FU0)\\
345 & 13.0 & 4.9 & 1235.24532 & -0.544 & 263.166 & -27.175 & (FU1) \\
347 & 13.0 & 4.9 & 1308.95261 & -3.621 & 264.053 & -25.778 & (FU0)\\
348 & 13.0 & 4.9 & 1324.98925 & -27.563 & 267.805 & -28.168 & (FU0)\\
349 & 13.0 & 4.9 & 1272.22405 & -27.694 & 264.174 & -27.709 & \\
351 & 13.0 & 4.9 & 1334.41973 & -22.024 & 262.929 & -28.009 & (FU0)\\

\hline
\end{tabular}
\caption[Outliers that passed detection pipeline]{Outliers in frequency range 1200-1400\,Hz that passed the detection pipeline excluding regions heavily contaminated with violin modes. Outliers marked with ``line'' had strong narrowband disturbances identified near the outlier location. We have marked outliers not consistent with the target signals at one of the semi-coherent \Fstat ~ follow-ups with ``(FU0/1/2)", depending on the stage at which they did not pass the detection thresholds.
Frequencies are converted to epoch GPS $1130529362$.}
\label{tab:PowerFluxOutliers3}
\end{center}
\end{table*}

\begin{table*}[htbp]
\begin{center}
\scriptsize
\begin{tabular}{D{.}{.}{2}D{.}{.}{2}cD{.}{.}{5}D{.}{.}{4}D{.}{.}{4}D{.}{.}{4}l}\hline
\multicolumn{1}{c}{Idx} & \multicolumn{1}{c}{SNR}  & \multicolumn{1}{c}{$\log_{10}\left({\textrm{GFA}}\right)$} &  \multicolumn{1}{c}{Frequency} & \multicolumn{1}{c}{Spindown} &  \multicolumn{1}{c}{$\RAJ$}  & \multicolumn{1}{c}{$\DECJ$} & Description \\
\multicolumn{1}{c}{}	&  \multicolumn{1}{c}{}	&  \multicolumn{1}{c}{}	& \multicolumn{1}{c}{Hz}	&  \multicolumn{1}{c}{nHz/s} & \multicolumn{1}{c}{degrees} & \multicolumn{1}{c}{degrees} & \\
\hline \hline
2 & 27.8 & -26.2 & 1457.98771 & -5.923 & 267.265 & -23.236 & Broad line in L1 \\
3 & 27.5 & -25.5 & 1495.87736 & -19.359 & 263.979 & -25.411 & Broad line in L1 \\
12 & 21.6 & -12.6 & 1469.41404 & -12.041 & 267.020 & -24.399 & Strong bin-centered line in L1 at 1469.3 Hz \\
15 & 20.9 & -11.0 & 1467.55257 & -19.926 & 265.951 & -29.741 & Broad disturbance in H1 (?) \\
18 & 20.2 & -9.6 & 1469.44100 & -1.796 & 269.546 & -24.514 & Strong bin-centered line in L1 at 1469.3 Hz \\
21 & 19.1 & -7.1 & 1421.11633 & -6.941 & 263.048 & -26.468 & Strong bin-centered line in L1 at 1421 Hz \\
23 & 18.1 & -5.1 & 1478.75395 & -0.269 & 268.825 & -26.084 & Broad line in H1 \\
26 & 17.8 & -4.4 & 1469.40683 & -14.406 & 263.178 & -26.341 & Strong bin-centered line in L1 at 1469.3 Hz \\
29 & 17.7 & -4.3 & 1421.11841 & -6.437 & 264.313 & -27.030 & Strong bin-centered line in L1 at 1421 Hz \\
30 & 17.6 & -4.2 & 1408.10431 & -25.129 & 266.189 & -23.920 & Strong bin-centered line in H1 at 1408 Hz \\
35 & 17.4 & -3.6 & 1418.20292 & -20.386 & 267.609 & -24.774 & Strong bin-centered line in L1 at 1418.1 Hz \\
36 & 17.1 & -3.1 & 1421.11841 & -6.436 & 264.313 & -27.043 & Strong bin-centered line in L1 at 1421 Hz \\
41 & 16.8 & -2.5 & 1484.50512 & -6.777 & 269.295 & -26.268 & Broad line in H1 \\
53 & 16.2 & -1.2 & 1467.47364 & -1.878 & 268.521 & -27.384 & Broad line in H1 \\
60 & 15.5 & 0.2 & 1478.75226 & -0.909 & 267.825 & -27.981 & Broad line in H1 \\
62 & 15.4 & 0.3 & 1499.44086 & -7.311 & 263.883 & -25.493 & Nearby broad lines in H1 and L1 \\
73 & 15.0 & 1.1 & 1484.75642 & -29.816 & 265.225 & -26.457 & Broad line in H1 \\
74 & 15.0 & 1.1 & 1401.69854 & -22.076 & 264.511 & -26.586 & (FU1) \\
88 & 14.7 & 1.7 & 1458.95808 & -15.962 & 268.240 & -26.171 & \\
90 & 14.7 & 1.8 & 1400.73383 & -8.001 & 268.040 & -25.138 & (FU1)\\
92 & 14.6 & 1.8 & 1492.34125 & -8.216 & 266.388 & -23.361 & Broad line in L1 \\
94 & 14.6 & 1.9 & 1497.84101 & -4.821 & 263.660 & -28.526 & Broad line in L1 \\
100 & 14.5 & 2.1 & 1484.48710 & -0.763 & 267.286 & -27.112 & Broad lines in H1 \\
132 & 14.0 & 3.1 & 1443.81841 & -9.151 & 268.034 & -25.533 & Strong bin-centered line in L1 at 1443.7 Hz \\
136 & 14.0 & 3.1 & 1497.84102 & -4.823 & 263.665 & -28.451 & Broad line in L1 \\
154 & 13.8 & 3.4 & 1498.67420 & -19.913 & 264.048 & -23.412 & Broad line in H1 \\
167 & 13.7 & 3.5 & 1472.02303 & -12.624 & 263.554 & -25.278 & Broad line in L1 \\
183 & 13.6 & 3.7 & 1402.87061 & -20.561 & 269.872 & -26.485 & \\
193 & 13.6 & 3.8 & 1457.63610 & -4.906 & 263.216 & -24.295 & Broad line in L1 \\
196 & 13.6 & 3.9 & 1454.89391 & -20.218 & 268.541 & -25.625 & \\
198 & 13.5 & 3.9 & 1442.61715 & -16.934 & 267.499 & -25.456 & \\
200 & 13.5 & 3.9 & 1462.09124 & -28.426 & 264.860 & -27.360 & Broad line in H1 \\
201 & 13.5 & 3.9 & 1488.98202 & -9.992 & 265.988 & -27.058 & (FU0)\\
204 & 13.5 & 3.9 & 1443.50058 & -8.113 & 268.399 & -26.734 & (FU0)\\
205 & 13.5 & 4.0 & 1496.13532 & -8.503 & 264.346 & -26.994 & Broad line in L1 \\
209 & 13.5 & 4.0 & 1499.43021 & -15.954 & 264.590 & -25.146 & Nearby broad lines in H1 and L1 \\
216 & 13.5 & 4.1 & 1498.67734 & -21.961 & 267.824 & -29.502 & Broad line in H1\\
224 & 13.4 & 4.2 & 1499.42849 & -16.246 & 263.688 & -28.354 & Nearby broad lines in H1 and L1 \\
226 & 13.4 & 4.2 & 1489.97625 & -29.213 & 269.022 & -24.757 & (FU0)\\
230 & 13.4 & 4.2 & 1465.36884 & -1.351 & 266.338 & -23.042 & (FU0)\\
231 & 13.4 & 4.3 & 1408.12849 & -2.961 & 264.812 & -28.640 & Strong bin-centered line in H1 at 1408 Hz \\
235 & 13.3 & 4.3 & 1482.00682 & -4.574 & 263.839 & -26.682 & Disturbed spectrum in H1 \\
239 & 13.3 & 4.3 & 1408.10786 & -17.896 & 270.286 & -26.385 & Strong bin-centered line in H1 at 1408 Hz \\
250 & 13.3 & 4.4 & 1459.40166 & -29.472 & 269.924 & -27.156 & (FU0)\\
252 & 13.3 & 4.4 & 1460.13953 & -25.686 & 267.954 & -26.966 & (FU0)\\
256 & 13.3 & 4.4 & 1408.11002 & -10.656 & 268.386 & -27.919 & Strong bin-centered line in H1 at 1408 Hz \\
259 & 13.3 & 4.4 & 1482.62319 & -22.633 & 264.730 & -27.878 & Nearby broad line in H1 \\
268 & 13.3 & 4.5 & 1445.84829 & -16.384 & 268.512 & -28.667 & (FU0)\\
275 & 13.2 & 4.5 & 1474.29669 & -28.242 & 265.215 & -26.055 & Nearby broad line in H1, disturbed H1 spectrum \\
277 & 13.2 & 4.5 & 1462.20888 & -19.916 & 268.963 & -25.106 & (FU0)\\
296 & 13.2 & 4.6 & 1478.10769 & -26.661 & 265.631 & -22.822 & Bin-centered line in L1 at 1478 Hz, disturbed H1 spectrum \\
301 & 13.2 & 4.6 & 1459.94553 & -3.364 & 265.599 & -29.328 &  (FU0)\\
302 & 13.2 & 4.6 & 1455.78804 & -1.216 & 263.403 & -28.173 & (FU1) \\
312 & 13.1 & 4.7 & 1472.96626 & -6.889 & 263.067 & -25.410 & (FU1)  \\
320 & 13.1 & 4.8 & 1432.48786 & -23.659 & 267.769 & -28.544 & (FU0)\\
321 & 13.1 & 4.8 & 1496.76907 & -17.039 & 262.628 & -26.112 & (FU0)\\
324 & 13.1 & 4.8 & 1499.42798 & -18.013 & 268.803 & -25.498 & Nearby broad lines in H1 and L1 \\
326 & 13.1 & 4.8 & 1475.08910 & -24.164 & 268.123 & -29.057 & Nearby strong line in H1, disturbed spectrum \\
328 & 13.1 & 4.8 & 1430.12480 & -23.206 & 266.342 & -24.490 & \\
329 & 13.1 & 4.8 & 1430.12480 & -23.204 & 266.342 & -24.625 & (FU0)\\
346 & 13.0 & 4.9 & 1423.41985 & -2.116 & 264.904 & -26.065 & (FU0)\\
352 & 13.0 & 4.9 & 1499.10849 & -0.374 & 267.745 & -26.798 & Broad line in L1 \\

\hline
\end{tabular}
\caption[Outliers that passed detection pipeline]{Outliers above 1400\,Hz that passed the detection pipeline excluding regions heavily contaminated with violin modes. Outliers marked with ``line'' had strong narrowband disturbances identified near the outlier location. We have marked outliers not consistent with the target signals at one of the semi-coherent \Fstat ~ follow-ups with ``(FU0/1/2)", depending on the stage at which they did not pass the detection thresholds.
Frequencies are converted to epoch GPS $1130529362$.}
\label{tab:PowerFluxOutliers4}
\end{center}
\end{table*}

\begin{table*}[htbp]
\begin{center}
\scriptsize
\begin{tabular}{D{.}{.}{2}D{.}{.}{2}cD{.}{.}{5}D{.}{.}{4}D{.}{.}{4}D{.}{.}{4}l}\hline
\multicolumn{1}{c}{Idx} & \multicolumn{1}{c}{SNR}  & \multicolumn{1}{c}{$\log_{10}\left({\textrm{GFA}}\right)$} &  \multicolumn{1}{c}{Frequency} & \multicolumn{1}{c}{Spindown} &  \multicolumn{1}{c}{$\RAJ$}  & \multicolumn{1}{c}{$\DECJ$}  \\
\multicolumn{1}{c}{}	&  \multicolumn{1}{c}{}	&  \multicolumn{1}{c}{}	& \multicolumn{1}{c}{Hz}	&  \multicolumn{1}{c}{nHz/s} & \multicolumn{1}{c}{degrees} & \multicolumn{1}{c}{degrees} \\
\hline \hline
4 & 27.3 & -25.0 & 508.26088 & -9.104 & 269.109 & -29.202 & Broad line in H1 at 508.222 \\
7 & 23.6 & -16.9 & 1030.75807 & -24.288 & 265.630 & -28.137 & Forest of strong lines in L1 \\
9 & 23.2 & -16.0 & 501.54872 & -22.486 & 264.981 & -23.309 & Large line in H1, violin mode region \\
10 & 22.7 & -14.9 & 1018.71452 & -13.869 & 268.055 & -23.245 & Strong line in L1 \\
11 & 22.0 & -13.4 & 505.62039 & -6.776 & 264.643 & -27.775 & Large lines in H1, violin mode region \\
14 & 21.0 & -11.2 & 1027.44609 & -28.797 & 266.018 & -29.859 & Forest of strong lines in L1\\
22 & 18.9 & -6.7 & 1014.13265 & -5.347 & 269.223 & -28.565 & Forest of strong lines in L1 \\
25 & 17.9 & -4.7 & 1030.76062 & -27.591 & 269.921 & -27.663 & Forest of strong lines in L1 \\
27 & 17.7 & -4.4 & 505.63342 & -0.324 & 266.622 & -29.454 & Large lines in H1, violin mode region \\
28 & 17.7 & -4.3 & 505.68386 & -20.164 & 267.657 & -25.001 & Large lines in H1, violin mode region\\
31 & 17.6 & -4.1 & 1008.58625 & 0.313 & 263.517 & -27.178 & Strong broad line in H1, line in L1 \\
38 & 17.0 & -2.9 & 505.72151 & -16.584 & 265.219 & -27.670 & Large line in H1, violin mode region \\
46 & 16.5 & -1.9 & 1006.00395 & -14.306 & 267.233 & -27.920 & Strong broad lines in H1 \\
48 & 16.5 & -1.9 & 1021.20375 & -22.584 & 267.795 & -24.931 & Lines in L1 \\
49 & 16.4 & -1.7 & 509.19731 & 0.039 & 266.911 & -29.803 & Violin mode region \\
50 & 16.3 & -1.5 & 1031.08895 & -27.254 & 263.465 & -24.793 & Lines in L1 \\
51 & 16.3 & -1.5 & 506.97395 & -12.404 & 265.161 & -27.594 & Large line in H1, violin mode region \\
52 & 16.3 & -1.4 & 1027.53960 & -15.576 & 269.243 & -24.104 & Forest of strong lines in L1 \\
54 & 15.8 & -0.6 & 509.19626 & -0.289 & 264.906 & -26.767 & Violin mode region \\
57 & 15.6 & -0.2 & 1017.17041 & -15.632 & 263.374 & -24.850 & \\
61 & 15.4 & 0.3 & 1029.13774 & -21.237 & 263.826 & -28.459 & Forest of strong lines in L1 (FU0)\\
64 & 15.3 & 0.5 & 505.72844 & -13.179 & 267.146 & -23.848 & Large broad and narrow lines in H1, L1, violin mode region \\
65 & 15.3 & 0.5 & 1027.53447 & -15.899 & 264.371 & -27.442 & Forest of strong lines in L1 \\
66 & 15.3 & 0.6 & 1014.13550 & -0.511 & 269.734 & -28.397 & Forest of strong lines in L1 \\
71 & 15.1 & 0.9 & 992.02121 & -22.994 & 269.830 & -28.298 & Strong broad line in H1, lines in L1 \\
79 & 14.9 & 1.4 & 503.01053 & -14.412 & 264.576 & -29.521 & Large lines in H1, violin mode region\\
83 & 14.8 & 1.5 & 1006.18012 & -21.694 & 263.313 & -24.036 & Strong bin-centered line in L1 at 1006.1 Hz \\
87 & 14.7 & 1.6 & 1030.76044 & -17.694 & 264.609 & -25.125 & Forest of strong lines in L1 \\
95 & 14.6 & 1.9 & 1003.74867 & -4.776 & 265.096 & -24.430 & Strong broad line in H1 \\
97 & 14.6 & 2.0 & 1013.03947 & -0.404 & 268.508 & -28.026 & Disturbed background in H1 \\
98 & 14.5 & 2.0 & 1026.14462 & -28.021 & 267.878 & -27.584 & Broad line in L1 \\
105 & 14.4 & 2.3 & 1004.02592 & -2.816 & 265.980 & -25.168 & (FU0)\\
108 & 14.3 & 2.4 & 1004.02591 & -2.814 & 265.980 & -25.265 & (FU0)\\
109 & 14.3 & 2.4 & 1003.73614 & -16.516 & 266.800 & -28.487 & Strong broad line in H1\\
110 & 14.3 & 2.5 & 1029.19672 & 1.012 & 270.018 & -27.237 & Forest of lines in L1\\
122 & 14.1 & 2.8 & 993.55783 & -25.844 & 265.053 & -24.977 & (FU0)\\
124 & 14.1 & 2.9 & 511.99612 & -0.903 & 266.165 & -25.709 & Sharp bin-centered line at 512 Hz\\
131 & 14.0 & 3.0 & 1000.06619 & -13.539 & 266.692 & -26.396 & Lines in H1 and L1\\
134 & 14.0 & 3.1 & 1011.00509 & -18.583 & 269.680 & -25.059 & Strong line in L1, disturbed H1 spectrum\\
153 & 13.8 & 3.4 & 1003.75296 & -3.369 & 265.458 & -27.670 & Strong broad line in H1\\
155 & 13.8 & 3.4 & 1025.63609 & -7.339 & 265.091 & -23.025 & Forest of lines in L1 \\
157 & 13.8 & 3.4 & 994.50537 & -0.149 & 264.616 & -28.649 & (FU0)\\
164 & 13.8 & 3.5 & 990.28898 & -0.769 & 264.972 & -26.915 & Disturbed H1 spectrum \\
168 & 13.7 & 3.5 & 1031.39556 & -7.501 & 263.845 & -23.332 & Forest of lines in L1 \\
178 & 13.7 & 3.7 & 1010.61505 & -17.337 & 265.335 & -24.634 & Disturbed H1 spectrum, lines in L1 \\
184 & 13.6 & 3.8 & 991.26021 & -23.166 & 266.511 & -27.602 & Disturbed H1 spectrum, lines \\
185 & 13.6 & 3.8 & 1006.00396 & -14.314 & 267.225 & -27.634 & Strong broad lines in H1, lines in L1 \\
187 & 13.6 & 3.8 & 1031.16138 & -14.838 & 264.511 & -23.556 & Forest of lines in L1 \\
197 & 13.6 & 3.9 & 509.19895 & 0.414 & 269.633 & -27.167 & Large lines, violin mode region \\
212 & 13.5 & 4.0 & 1017.38348 & -16.619 & 268.845 & -24.206 & (FU0)\\
217 & 13.5 & 4.1 & 1018.82920 & -17.391 & 264.399 & -26.832 & (FU0)\\
221 & 13.4 & 4.1 & 1011.02816 & -16.411 & 267.760 & -22.824 & Disturbed H1 spectrum, lines in L1 \\
222 & 13.4 & 4.1 & 1016.77536 & -0.176 & 267.870 & -23.024 & (FU0)\\
225 & 13.4 & 4.2 & 993.08901 & -25.739 & 269.788 & -25.063 & (FU1) \\
234 & 13.4 & 4.3 & 992.25622 & -22.391 & 265.644 & -27.798 & Forest of lines in H1 and L1 \\
240 & 13.3 & 4.3 & 1005.02179 & -22.107 & 269.152 & -28.822 & (FU0)\\
280 & 13.2 & 4.5 & 1031.49141 & -25.294 & 266.685 & -27.054 & Forest of lines in L1 \\
281 & 13.2 & 4.5 & 1013.13666 & -22.974 & 268.774 & -25.644 & Large disturbance in H1 \\
292 & 13.2 & 4.6 & 997.43474 & -5.569 & 266.898 & -26.932 & Strong bin-centered line in L1 at 997.4, disturbed H1 spectrum \\
304 & 13.2 & 4.6 & 1002.78826 & -1.404 & 268.608 & -27.280 & Strong line in L1 \\
306 & 13.1 & 4.7 & 1000.06213 & -15.341 & 264.654 & -27.047 & Lines in H1 and L1 (FU0)\\
315 & 13.1 & 4.7 & 991.62069 & -12.746 & 263.013 & -27.125 & Strong broad line in L1 \\

\hline
\end{tabular}
\caption[Outliers near violin modes that passed detection pipeline]{Outliers in 495-520\,Hz and 990-1033\,Hz regions heavily contaminated with violin modes. Outliers marked with ``line'' had strong narrowband disturbances identified near the outlier location. We have marked outliers not consistent with the target signals at one of the semi-coherent \Fstat ~ follow-ups with ``(FU0/1/2)", depending on the stage at which they did not pass the detection thresholds.
Frequencies are converted to epoch GPS $1130529362$.}
\label{tab:PowerFluxOutliersViolin}
\end{center}
\end{table*}


\begin{thebibliography}{99}

\bibitem{rilesmpla} Recent searches for continuous gravitational waves, K.~Riles, Mod.\ Phys.\ Lett.\ A {\bf 32} 1730035 (2017).


\bibitem{O1EH} First low-frequency Einstein@Home all-sky search for continuous gravitational waves in Advanced LIGO data, B.~P.~Abbott \etal\ (LIGO Scientific Collaboration and Virgo Collaboration), Phys.\ Rev.\ D {\bf 96} 122004 (2017)

\bibitem{O1AllSky1}
All-sky search for periodic gravitational waves in the O1 LIGO data,
B.~P.~Abbott \etal\ (LIGO Scientific Collaboration and Virgo Collaboration), Phys.\ Rev.\ D {\bf 96} 062002 (2017).

\bibitem{O1AllSky2}
Full Band All-sky Search for Periodic Gravitational Waves in the O1 LIGO Data,
B.~P.~Abbott \etal\ (LIGO Scientific Collaboration and Virgo Collaboration), Phys.\ Rev.\ D {\bf 97} 102003 (2018).

\bibitem{o1_allsky3} First loosely coherent all-sky search for periodic gravitational waves in the O1 LIGO data, V.~Dergachev, M.A.~Papa, \url{https://arxiv.org/abs/1902.05530}, submitted to PRL.

\bibitem{orionspur}
A search of the Orion spur for continuous gravitational waves using a "loosely coherent" algorithm on data from LIGO interferometers
J.~Aasi \etal\ (LIGO Scientific Collaboration and Virgo Collaboration), Phys.\ Rev.\ D {\bf 93}, 042006 (2016).


\bibitem{losc} M.~Vallisneri \etal\ "The LIGO Open Science Center", proceedings of the 10th LISA Symposium, University of Florida, Gainesville, May 18-23, 2014, arxiv:1410.4839.

\bibitem{o1_data} https://doi.org/10.7935/CA75-FM95LIGO Open Science Center, \url{https://losc.ligo.org}.

\bibitem{aligo} Advanced LIGO, J.~Aasi \etal\ (LIGO Scientific Collaboration), Class.\ Quantum Grav.\  {\bf 32} 7 (2015) .

\bibitem{O1DetectorPaper}
  GW150914: The Advanced LIGO Detectors in the Era of First Discoveries,
  B.~P.~Abbott \etal\ (LIGO Scientific Collaboration and Virgo Collaboration), Phys.\ Rev.\ Lett. {\bf 116} 131103 (2016).

\bibitem{loosely_coherent}
On blind searches for noise dominated signals: a loosely coherent approach,
V.~Dergachev, Class.\ Quantum Grav.\ {\bf 27}, 205017 (2010).

\bibitem{loosely_coherent2}
Loosely coherent searches for sets of well-modeled signals,
V.~Dergachev,
Phys.\ Rev.\ D {\bf 85}, 062003 (2012).

\bibitem{loosely_coherent3} Efficient loosely coherent searches for medium scale coherence lengths, V.~Dergachev, \url{https://arxiv.org/abs/1807.02351}, submitted to PRD.

\bibitem{McDanielJohnsonOwen} Maximum elastic deformations of relativistic stars, N.K.~Johnson-McDaniel and B.J.~Owen, Phys.\ Rev.\ D {\bf 88} 044004 (2013).

\bibitem{Covasetal}
Identification and mitigation of narrow spectral artifacts that degrade searches for persistent gravitational waves in the first two observing runs of Advanced LIGO,
P.~Covas \etal, Phys.\ Rev.\ D {\bf 97} 082002 (2018).

\bibitem{data} See EPAPS Document No. [number will be inserted by
publisher] for numerical values of upper limits.

\bibitem{S4IncoherentPaper}
All-sky search for periodic gravitational waves in LIGO S4 data,
B.~Abbott \etal\ (LIGO Scientific Collaboration),
Phys.\ Rev.\ D \textbf{77}, 022001 (2008).

\bibitem{EarlyS5Paper}
All-sky LIGO Search for Periodic Gravitational Waves in the Early S5 Data,
 B.~P.~Abbott \etal\ (LIGO Scientific Collaboration), Phys.\ Rev.\ Lett.\ \textbf{102}, 111102 (2009).

 \bibitem{FullS5Semicoherent}
All-sky Search for Periodic Gravitational Waves in the Full S5 Data,
B.~Abbott \etal ~ (The LIGO and Virgo Scientific Collaboration),  {\it Phys. Rev. } {\bf D 85}, 022001 (2012).


\bibitem{universal_statistics}
A Novel Universal Statistic for Computing Upper Limits in Ill-behaved Background,
V.~Dergachev,
Phys.\ Rev.\ D \textbf{87}, 062001 (2013).

\bibitem{TerzanPulsarPaper1}
Radio Pulsars in Terzan 5,
A.G.~Lyne \etal, M.N.R.A.S. {\bf 316} 491 (2000). 

\bibitem{TerzanPulsarPaper2}
Discovery of Three New Millisecond Pulsars in Terzan 5,
M. Cadelano \etal, to appear in Astroph. J, arXiv:1801.09929, January 2018.  
  
\bibitem{ATNF} R.N.~Manchester, G.B.~Hobbs, A.~Teoh, M.~Hobbs, AJ, 129, 1993-2006 (2005)

\bibitem{GW150914} 
  B.~P.~Abbott {\it et al.} [LIGO Scientific and Virgo Collaborations],
  Phys.\ Rev.\ Lett.\  {\bf 116}, no. 6, 061102 (2016)
  doi:10.1103/PhysRevLett.116.061102
  [arXiv:1602.03837 [gr-qc]].

\bibitem{SensDepthDef} 
  B.~Behnke, M.~A.~Papa and R.~Prix,
  Postprocessing methods used in the search for continuous gravitational-wave signals from the Galactic Center,
  Phys.\ Rev.\ D {\bf 91}, no. 6, 064007 (2015)

\bibitem{Dreissigacker:2018afk} 
  C.~Dreissigacker, R.~Prix and K.~Wette,
  Fast and Accurate Sensitivity Estimation for Continuous-Gravitational-Wave Searches,
  Phys.\ Rev.\ D {\bf 98}, no. 8, 084058 (2018)


\bibitem{LVCRadiometerO1} 
  B.~P.~Abbott {\it et al.} [LIGO Scientific and Virgo Collaborations],
  Directional Limits on Persistent Gravitational Waves from Advanced LIGO First Observing Run,
  Phys.\ Rev.\ Lett.\  {\bf 118}, no. 12, 121102 (2017)

\bibitem{O2AllSkyLVC}
B.~P.~Abbott {\it et al.} [LIGO Scientific and Virgo Collaborations],
 All-sky search for continous gravitational waves from isolated neutron stars using Advanced LIGO O2 data,
 arXiv:1903:01901



\end{thebibliography}
\end{document}